\documentclass[aip,amsmath,amssymb,jcp]{revtex4-1}
\usepackage {CJK}

\usepackage[utf8]{inputenc}
\usepackage[T1]{fontenc}
\usepackage{graphicx}

\usepackage{xcolor}

\usepackage{float}
\makeatletter
\let\newfloat\newfloat@ltx
\makeatother

\newcommand{\diff}[2]{\frac{\partial#1}{\partial #2}}

\def\bb{\boldsymbol{b}}

\def\xb{\boldsymbol{x}}

\def\Xb{\boldsymbol{X}}
\def\Xbu{\boldsymbol{X}^u}
\def\Wb{\boldsymbol{W}}

\def\RR{\mathbb{R}}  
\def\EE{\mathbb{E}}

\def\<{\langle} \def\>{\rangle}

\def\<{\langle} \def\>{\rangle}
\newcommand{\rr}{\pmb{r}}

\newcommand{\dPP}{\mathrm{d}\mathbb{P}}

\def\Xb{\boldsymbol{X}}
\def\Xbu{\boldsymbol{X}^u}
\def\Wb{\boldsymbol{W}}

\draft 

\begin{document}


\title[Scalable GNNs for dynamical fluctuations]{Physics-informed graph neural networks enhance scalability of variational nonequilibrium optimal control}


\author{Jiawei Yan (闫嘉伟)}
\email{yanjw@stanford.edu}
\affiliation{Department of Chemistry, Stanford University, Stanford, CA 94305, USA}
\author{Grant M. Rotskoff}
\email{rotskoff@stanford.edu}
\affiliation{Department of Chemistry, Stanford University, Stanford, CA 94305, USA}

\date{\today}

\begin{abstract}%
 When a physical system is driven away from equilibrium, the statistical distribution of its dynamical trajectories informs many of its physical properties.
 Characterizing the nature of the distribution of dynamical observables, such as a current or entropy production rate, has become a central problem in nonequilibrium statistical mechanics.
 Asymptotically, for a broad class of observables, the distribution of a given observable satisfies a large deviation principle when the dynamics is Markovian, meaning that fluctuations can be characterized in the long-time limit by computing a scaled cumulant generating function.
 Calculating this function is not tractable analytically (nor often numerically) for complex, interacting systems, so the development of robust numerical techniques to carry out this computation is needed to probe the properties of nonequilibrium materials.
 Here, we describe an algorithm that recasts this task as an optimal control problem that can be solved variationally.
 We solve for optimal control forces using neural network ans\"atze that are tailored to the physical systems to which the forces are applied.
 We demonstrate that this approach leads to transferable and accurate solutions in two systems featuring large numbers of interacting particles.
\end{abstract}

\begin{CJK*}{UTF8}{}
\CJKfamily{gbsn}
\maketitle
\end{CJK*}

%

\section{Introduction}

At equilibrium, a physical system relaxes to a time-invariant distribution with a statistical weight that can, in many cases, be computed easily.
Straightforward characterization of the relative probabilities of distinct states facilitates the use of Markov Chain Monte Carlo algorithms, and, as a result, expectations of observables are tractable to compute in many cases.
When we drive a system away from equilibrium, either through the application of an external perturbation or with an energy-consuming dynamics that inherently breaks detailed balance, we lose the luxury of this facile description and knowledge of the universal distribution of states.
While it is possible in some cases to characterize steady-state distributions of complex, interacting, nonequilibrium systems~\cite{toner1995long, toner1998flocks, lazarescu_exact_2015, Martin_2021_statistical, fodor2021irreversibility, cates_motilityinduced_2015}, as an alternative, we adopt a manifestly \emph{dynamical} point of view.

The large deviation perspective on nonequilibrium statistical mechanics focuses less on static states of a system and more on time-dependent trajectories~\cite{garrahan_dynamical_2007, touchette_large_2009}.
Ruelle developed a thermodynamic formalism~\cite{ruelle_thermodynamic_2009} based on Donsker-Varadhan large deviation theory that has become a widely-used framework to understand the statistical fluctuations of dynamical behavior, from glassiness~\cite{garrahan_dynamical_2007, garrahan_firstorder_2009, hedges_dynamic_2009}, to dynamical phase transitions~\cite{gingrich_heterogeneity_2014, grandpre_entropy_2021}, to thermodynamic uncertainty relations~\cite{barato_thermodynamic_2015, gingrich_dissipation_2016}.
The central object of study within this theory is the scaled cumulant generating function (SCGF) for a dynamical observable $A_T.$
This object is defined as the limit of an expectation over trajectories,
\begin{equation}
 \psi(\lambda) = \lim_{T \to\infty} \frac{1}{T} \log \EE_{\Xb} e^{\lambda T A_T},
 \label{eq:scgf1}
\end{equation}
plays a role akin to that of a free energy for an equilibrium system.
The Legendre-Fenchel transform of $\psi(\lambda)$ defines the \emph{large deviation rate function} $I(a)$, which describes the rate of decay of the probability that $A_T \in [a,a+da]$
\begin{equation}
 P(A_T \in [a, a+da]) \asymp e^{-T I (a)} da.
 \label{eq:ldf}
\end{equation}
These two objects contain identical statistical information in the long-time limit, but both \eqref{eq:scgf1} and \eqref{eq:ldf} are notoriously difficult to compute because it involves exponential averages that are dominated by rare events.

Here, we outline a robust and transferable strategy for estimating \eqref{eq:scgf1} that applies to two broad classes of interacting models: interacting particles systems with diffusive dynamics and Markov jump processes where local interactions dictate jump rates.
In both cases, we construct a neural network ansatz for \emph{virtual} control forces that realize the rare trajectories necessary to compute~\eqref{eq:scgf1}.
Our results demonstrate that carefully incorporating physical symmetries into the architecture of the neural network ansatz leads to an accurate, straightforward algorithm that can be deployed on a wide variety of systems.

The present work builds on recent developments using machine learning~\cite{yan2022learning,das_reinforcement_2021,Casert_Dynamical_2021}, that provide a generic framework for learning optimal control forces to minimize a variational objective for~\eqref{eq:scgf1}.
These control forces guide the dynamics to realize the statistically rare fluctuations that lead to the corresponding rare values of the observable $A_T$.
Here, we demonstrate that this algorithm can be specifically tailored for interacting systems with appropriate inductive biases.
Physical constraints, such as translational invariance, are built into the neural network to achieve fast and transferable training.
The resulting control forces efficiently define a many-body control force to achieve rare fluctuations.
One key advantage to the approach taken here is that, unlike existing approaches, the new neural network architecture we employ allows \emph{transfer training}, in which a small, easy-to-simulate system is used to generate training data, but the networks are subsequently deployed on larger systems.

\subsection{Existing numerical approaches}
Many numerical algorithms have been proposed to access the rare dynamical fluctuations that determine a large deviation rate function.
These algorithms vary in their approach; those based on importance sampling~\cite{bucklew2004, ray_exact_2018, ferre_adaptive_2018} sample a dynamical process distinct from the original process, but re-weight those samples accordingly.
On the other hand, cloning algorithms~\cite{grassberger2002, giardina_direct_2006, lecomte_numerical_2007, ray_constructing_2020} implement a population dynamics at the level of trajectories---trajectories are killed and cloned with to re-weight expectation values with the exponential statistical bias in the scaled cumulant generating function.
State-of-the-art methods employ a combination of cloning and adaptive sampling~\cite{nemoto_populationdynamics_2016} using parameter dependent potential functions to guide the birth-death dynamics.
Along these lines, a series of papers has employed genetic algorithms~\cite{jacobson_direct_2019, whitelam_evolutionary_2020, whitelam_learning_2020,Casert_Dynamical_2021,Sprague_2021} and has shown that this approach can powerfully exploit the variational nature of the problem.

Numerically exact approaches are tractable for sufficiently small state spaces; the limit in~\eqref{eq:scgf1} can be computed by obtaining the maximum eigenvalue of a $\lambda$-dependent operator~\cite{lebowitz_gallavotti_1999}.
For some lattice models, this eigenvalue problem is therefore amenable to treatment using matrix product states (MPS) and numerical techniques based on density matrix renormalization group (DMRG) yield very accurate results when used with a tensor network ansatz~\cite{helms2019dynamical,helms2020dynamical, banuls2019using}.
Of course, the systems for which DMRG can be used are inherently limited and it remains challenging to extend these methods to models with continuous state spaces like interacting particle systems.

Machine learning approaches to this problem have relied for the most part on Malliavin weight sampling~\cite{das_variational_2019} or reinforcement learning~\cite{rose_reinforcement_2021, das_reinforcement_2021} to parameterize polynomial force functions that drive sampling.
Most closely related to the present work~\citet{Casert_Dynamical_2021, yan2022learning} use a neural network to represent a biasing force.

Our approach to estimating large deviation rate functions falls broadly under the umbrella of importance sampling for dynamical trajectories, a topic on which there has been substantial progress in recent years. Improvements in transition path sampling~\cite{das_direct_2022} and nonequilibrium sampling methods~\cite{rotskoff_dynamical_2019,thin_invertible_2021}.
Also related to the present work, machine learning algorithms dedicated to learning a force field or many-body potential in interacting particle systems have become a topic of intense study, including \cite{zhang_deep_2018,schutt_schnet_2018,wang_deeppmd_2018,frishman_2020_learning}.
Many of these works incorporate physical inductive biases that facilitate the effort to construct a force that preserves physical invariances and equivariances~\cite{zhang_deep_2018, satorras21a, schutt_schnet_2018}.

\section{Sampling as an optimal control problem}

In this paper we consider interacting particle systems described by a Markovian dynamics, which could either be a Markov jump process (cf. Appendix~\ref{sc:scgf_jump}) or a stochastic differential equation (SDE):
\begin{equation}
 \label{eq:sde}
 d\Xb_t = b(\Xb_t) dt + \sigma d\Wb_t,
\end{equation}
where $\Xb_t\in\RR^d$ is the state of the system, $b:\RR^d \to \RR^d$ is the drift function, and $\Wb_t$ is a Wiener process acting as a noise source, which is multiplied by a matrix $\sigma$.
For systems in equilibrium $b$ is the gradient of a potential energy function describing the interactions among the particles, but for systems out of equilibrium it may have non-conservative contributions.
For simplicity, we assume that $\sigma$ is independent of space and that the corresponding diffusion tensor $D= \sigma\sigma^T$ is invertible. Moreover, we assume that $\Xb_t$ is ergodic, which means that it has a unique stationary probability density, reached from any initial distribution in the long-time limit.

For nonequilibrium systems, fluctuations in time-extensive observables, such as currents, entropy production, and work yield information about dynamical response~\cite{maes_response_2020} and constraints on the dissipation rate~\cite{barato_thermodynamic_2015,gingrich_dissipation_2016}.
Following the now standard formulation~\cite{touchette_large_2009}, we consider a class of dynamical observables $A_T$ which has the following generic form:
\begin{equation}
 \label{eq:observable}
 A_T = \frac1T \int_0^T f(\Xb_t) dt + \frac1T \int_0^T g(\Xb_t) \circ d\Xb_t,
\end{equation}
where $f:\nobreak\RR^d\to \RR$ we call ``density-like'' and $g:\RR^d \to \RR^d$ we call ``current-like''.
Depending on the choices of these two functions, \eqref{eq:observable} can represent a large class of physical quantities, including the current, dynamical activity, work, and entropy production~\cite{touchette_introduction_2017}.
Computing the statistical distribution of $A_T$ is, of course, not analytically tractable in all but the simplest systems, however it often satisfies a large deviation principle:
\begin{equation}
 \label{eq:ldp}
 p(A_T = a) \asymp e^{-T I(a)}
\end{equation}
where $I(a)$ is called the ``rate function'' and the symbol $\asymp$ denotes equivalence up to logarithmic corrections.
The asymptotic expression \eqref{eq:ldp} can be regarded as a generalization of the central limit theorem---Gaussian fluctuations around the mean value lead to a quadratic expression for $I(a)$.
To compute $I(a)$ we can equivalently evaluate the Legendre-Fenchel transform of the scaled cumulant generating function (SCGF) $\psi(\lambda)$:
\begin{equation}
 \label{eq:scgf}
 \psi(\lambda) = \lim_{T\to\infty} \frac1T \log \EE_{\Xb} e^{\lambda T A_T},
\end{equation}
where $\EE_{\Xb}$ denotes an ensemble average over trajectories generated from \eqref{eq:sde}, and $\lambda$ is a real number.
When $\lambda\neq 0$, the average in \eqref{eq:scgf} is dominated by trajectories that realize rare values of the observable $A_T$, and, as a result, it is difficult to evaluate $\psi(\lambda)$ by directly simulating trajectories.

\subsection{Rigorous formulation of Active Importance Sampling}

Importance sampling can ameliorate the sampling problem required to compute~\eqref{eq:scgf} to some extent, but it is often difficult to use because one must determine an appropriate biased sampling distribution.
In the context of sampling dynamical trajectories, we must simulate the dynamics of an alternative SDE
\begin{equation}
 \label{eq:sde_u}
 d\Xbu_t = u_t(\Xbu_t) dt + \sigma d\Wb_t,
\end{equation}
and then the SCGF can be estimated simply by reweighting the average
\begin{equation}
 \psi(\lambda) = \lim_{T\to \infty} \frac{1}{T}\log \EE_{\Xbu} \left( e^{\lambda T A_T} \frac{\dPP[\Xbu]}{\dPP_u[\Xbu]} \right).
\end{equation}
Here $\EE_{\Xbu}$ denotes the fact that the expectation is taken over trajectories from \eqref{eq:sde_u} rather than the original dynamics \eqref{eq:sde}.

The main task here is to identify an optimal or near optimal control force $u_t(\Xbu_t)$ which can turn the rare event we aim to sample into a typical one. 
To achieve this goal, we adapt the weak convergence approach developed by \citet{dupuis_importance_2004} and extend it to continuous-time Markov processes. Let $\dPP_\nu$ be a path measure on $[0,T]$ such that its Radon-Nikodym derivative with respect to the path measure of the original dynamics is
\begin{equation}
 \label{eq:dp_nu}
 \frac{\dPP_\nu}{\dPP}(\omega) = \frac{e^{\lambda T A_T}}{\EE[e^{\lambda T A_T}]}.
\end{equation}
Then for any fixed $T$, the Kullback-Leibler (KL) divergence between two path measure $\dPP_u$ and $\dPP$ is
\begin{align}
 \frac{1}{T} \mathcal{D}_{KL}[\dPP_u \| \dPP]
 = & \frac{1}{T} \int \log\frac{\dPP_u}{\dPP}(\omega)\dPP_u(\omega), \nonumber                                                            \\
 = & \frac{1}{T} \int \log\frac{\dPP_u}{\dPP_\nu}(\omega)\dPP_u(\omega) + \int \log\frac{\dPP_\nu}{\dPP}(\omega)\dPP_u(\omega), \nonumber \\
 \label{eq:kl_3}
 = & \frac{1}{T}\mathcal{D}_{KL}[\dPP_u \| \dPP_\nu] + \lambda \EE_u[A_T] - \frac{1}{T}\log\EE[e^{\lambda T A_T} ].
\end{align}
Simply rearranging the expression \eqref{eq:kl_3}, we see that
\begin{equation}
 \frac{1}{T}\log\EE[e^{\lambda T A_T} ] -  \frac{1}{T}\mathcal{D}_{KL}[\dPP_u \| \dPP_\nu] = \lambda \EE_u[A_T] - \frac{1}{T} \mathcal{D}_{KL}[\dPP_u \| \dPP].
\end{equation}

Assuming all the limits exist, we take the limit $T\rightarrow\infty$ and then maximize over $u$ on both sides:
\begin{align}
   & \sup_{u} \left\{ \lim_{T\rightarrow\infty}\frac{1}{T}\log\EE[e^{\lambda T A_T} ] - \lim_{T\rightarrow\infty}\frac{1}{T}\mathcal{D}_{KL}[\dPP_u \| \dPP_\nu] \right\} \\
 = & \sup_{u} \left\{ \lim_{T\rightarrow\infty}\EE_u[A_T] - \lim_{T\rightarrow\infty}\frac{1}{T} \mathcal{D}_{KL}[\dPP_u \| \dPP] \right\}
\end{align}
Since the KL-divergence is always non-negative and the first term on the left is independent with $u$, the left side reaches its supremal value only when the KL-divergence term vanishes, therefore,
\begin{align}
 \label{eq:variational_formula}
 \psi(\lambda) = \lim_{T\rightarrow\infty}\frac{1}{T}\log\EE[e^{\lambda T A_T} ] = \sup_{u} \lim_{T\rightarrow\infty}\left\{ \EE_u[A_T] - \frac{1}{T} \mathcal{D}_{KL}[\dPP_u \| \dPP] \right\}.
\end{align}

The derivations above indicate that the optimal control force $u^*$ has the property
\begin{equation}
 \frac{1}{T}\mathcal{D}_{KL}[\dPP_{u^*} \| \dPP_\nu] = 0,
\end{equation}
so $\dPP_{u^*}$ is equivalent with the path measure $\dPP_\nu$ we construct in \eqref{eq:dp_nu}. However, the Radon-Nikodym derivative does not necessarily imply the explicit form of the controlled process \eqref{eq:sde_u}, nevertheless to mention that knowing the explicit form of \eqref{eq:dp_nu} requires knowing $\EE[e^{\lambda T A_T}]$, which is exactly we aim to compute. 
Therefore, we cannot direct construct a control force to match \eqref{eq:dp_nu} but we can optimize \eqref{eq:variational_formula} to determine the control force.

The KL-divergence in \eqref{eq:variational_formula} between path measure of the controlled dynamics \eqref{eq:sde_u} and the original dynamics \eqref{eq:sde} can be evaluated explicitly since \eqref{eq:sde_u} and \eqref{eq:sde} have the same noise tensor $\sigma$. 
According to Girsanov theorem, the Radon-Nikodym derivative is determined by the Martingale process~\cite{oksendal_stochastic_1992}
\begin{equation}
 \label{eq:girsanov_sde}
 \frac{\dPP_u}{\dPP} = \exp\left\{\int_0^T (u-b)\sigma^{-1} dW_t + \frac{1}{2} \int_0^T (u-b)^\top D^{-1}(u-b) dt \right\},
\end{equation}
and thus the stochastic integral term vanishes when taking the expectation over the path space.
Therefore,
\begin{align}
 \frac{1}{T} \mathcal{D}_{KL}[\dPP_u \| \dPP] = \frac{1}{2T} \EE_u\left[\int_0^T (u-b)^\top D^{-1}(u-b) dt\right].
\end{align}

\section{Building physically motivated representations of optimal control forces}
\label{sc:algo}

\begin{figure}
 \centering
 \includegraphics[width=.8\textwidth]{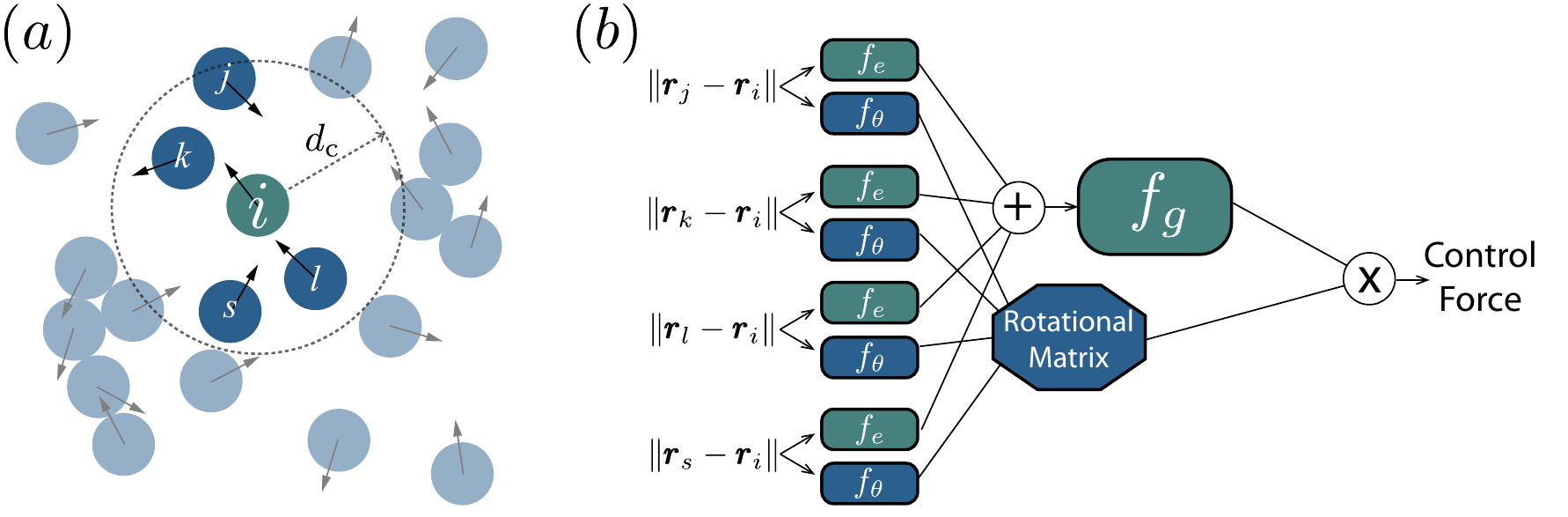}
 \caption{($a$) The information gathered by the graph neural network is limited to a particle's local environment, defined by a cutoff distance $d_{\rm c}$. ($b$) The graph neural network architecture, where $f_g$, $f_e$, and $f_\theta$ are three independent multilayer perceptrons (see \eqref{eq:nn}).}
 \label{fig:gnn_illustration}
\end{figure}

Computing SCGF through optimizing \eqref{eq:variational_formula} can be carried by representing the control force $u$ by a deep neural network. 
We have previously shown that this approach can obtain accurate results with significantly lower computational overhead than adaptive cloning methods~\cite{yan2022learning}. 
However, interacting particle systems pose a particular challenge because probing collective effects such as phase transitions require simulations with hundreds of particles. 
Though it is still possible to use one large deep neural network to represent the control force of the entire system, i.e., all particles, training becomes very difficult as the number of particle increases, and also the memory usage becomes extensive.
One solution to this problem is to leverage the fact that the control force for any given particle should depend most strongly on its local environment, i.e., the relative positions of other particles nearby, as illustrated in Fig.~\ref{fig:gnn_illustration} ($a$). 
In addition, the control force should also be restricted by physical laws and constraints, including translational invariance, permutation invariance, and equivariance. 
With these constraints, we here propose a representation of the control forces that is tailored for representing the control force for interacting particle systems.

As mentioned above, we assume that the control force $u$ depends only on one particle's local environment, and should be constrained by physical laws and invariance including:
\begin{itemize}
 \item Permutation invariance: the control force for a given particle remains the same when exchanging the index of its two neighboring particles
 \item Translational invariance: the control force remains the same when shifting the coordinates by a constant vector
 \item Equivariance: the direction of the control force rotates by the same angle as the coordinates rotate.
\end{itemize}

A general neural network, e.g., a multilayer perceptron (MLP), does not automatically satisfy the constraints outlined above, though it is possible to approximately realize these properties by augmenting the data to learn this physical information. 
A more natural approach is to incorporate a physical inductive bias directly into the architecture of the neural network, sometimes called physics-informed machine learning~\cite{karniadakis_physics_2021}.
We use a representation that satisfies all the three properties above, without sacrificing the expressiveness of deep neural networks.
The local environment of a given particle, defined by a cutoff distance $d_{\rm c}$, limits the spatial extent of interactions that determine the control force.
For particle $i$, the control force is a function of all the neighboring particles $j$ such that $|\rr_j-\rr_i| \leqslant d_{\rm c}$ (see Fig.~\ref{fig:gnn_illustration} for illustrations).

As described above, a generic neural network does not maintain equivariance of the forces, therefore we compute the direction and magnitude of the control force separately\cite{satorras21a}. Furthermore, the output of the neural network should be insensitive to any re-ordering of neighboring particles to ensure permutation invariance.
We use a message-passing graph neural network architecture, illustrated in Fig.~\ref{fig:gnn_illustration}(b), in which the messages
\begin{equation}
 m_{ij} = f_{\rm e}(\|\xb_j - \xb_i\|) \cdot \phi(\|\xb_j - \xb_i\|)
\end{equation}
are passed between the particle labelled $i$ and neighboring particles indexed by $j$. 
The function $\phi(x)$ is a smooth decay function in the range of $[0, d_{\rm c}]$ satisfying $\phi(0)=1$ and $\phi(d_{\rm c})=0$ since the control force should smoothly decay to $0$ as it reaches the cutoff distance. In the present work, we use $\phi(x) = \sqrt{[\cos(\pi x/d_{\rm c})+1]/2}$, though others functions may also function well.
The function $f_{\rm e}$ represents some neural network, here we also use an MLP. 
The magnitude of the control force is parameterized by an additional neural network which takes as an input the sum over messages,
\begin{equation}
 |\pmb{u}_i| = f_g(\sum_j m_{ij}) = f_g\left[ \sum_j f_e(\|\xb_j - \xb_i\|)\cdot \phi(\|\xb_j - \xb_i\|) \right].
\end{equation}
We include the sum over messages to maintain permutation invariance. 

A rotationally equivariant transformation can be implemented by parameterizing an orthogonal matrix that depends only on the inter-particle distance; we use
\begin{equation}
 \frac{\pmb{u}_i}{|\pmb{u}_i|} = \sum_j R\left[f_\theta(\|\xb_j - \xb_i\|)\right] \frac{\xb_j - \xb_i}{\|\xb_j - \xb_i\|},
\end{equation}
where $R(\cdot)$ is the rotation matrix
\begin{equation}
\label{eq:rotational_mat}
 R(\theta) =
 \begin{bmatrix}
  \cos(\theta) & -\sin(\theta) \\
  \sin(\theta) & \cos(\theta)
 \end{bmatrix}.
\end{equation}
Rotation matrices in two-dimensions are normal matrices, i.e., $R(\theta_1)R(\theta_2) = R(\theta_2)R(\theta_1)$ for any two $\theta_1$ and $\theta_2$, if we rotate the coordinate by an angle $\theta$, the direction of the control force will also rotate by an angle $\theta$. 
The message-passing framework we use is similar to \citet{satorras21a} but we include the rotation matrix \eqref{eq:rotational_mat} to allow more expressive representations of the direction of the control force.
Because we use the commutativity of rotation matrices in two dimensions, the representation we use will require further development to be extended to three-dimensional inputs. 
For those systems, how to construct a flexible representation of the direction of the control force while remaining equivariant is a challenge for future studies.

The total control force on particle $i$ can be represented by a combination of three neural networks $f_\theta$, $f_g$ and $f_e$,
\begin{equation}
 \label{eq:nn}
 u_i = \sum_j R[\phi_\theta(\|\xb_j - \xb_i\|)]\frac{\xb_j - \xb_i}{\|\xb_j - \xb_i\|}\cdot f_g\left[ \sum_j f_e(\|\xb_j - \xb_i\|) \right]
\end{equation}
where $f_\theta: \RR^1\rightarrow\RR^1$, $f_e: \RR^1\rightarrow\RR^h$ and $f_g: \RR^h\rightarrow\RR^2$ are three different neural networks composed independently by a stack of MLPs with linear residue followed by a linear layer to match the dimension.
That is, 
\begin{align}
 \label{eq:mlp_1}
 f_{g,e,\theta} & = P \circ L_1 \circ L_2 \circ \cdots \circ L_n \circ W           \\
 \label{eq:mlp_2}
 L_i(x)               & = \phi[W_{i,2}\cdot\phi(W_{i,1}\cdot x + b_{i,1}) + b_{i,2}] + x,
\end{align}
where $\phi$ is a nonlinear activation function. We use $\phi(x)=\max(0,x)$, the ReLU nonlinearity, through this work. 
The map $P$ is a padding layer that matches the input dimension with the hidden layer dimension. 
The input of $f_g$ is $0$ when there are no particles within the cutoff distance. 
Accordingly, we set the bias $b_{i,j}$ parameters to be $0$ for $f_g$ to ensure the control force is $0$ in this case.
We document further implementation details in the our code, which is available under an open-source license~\cite{code}.
The resulting representation of the control force has no information on the total number of particles, thus can be trained in a small system to quickly obtain the optimal control forces and scale up to large systems thereafter.

\section{Sampling dynamical fluctuations near phase transitions}

\subsection{Current fluctuations in the asymmetric simple exclusion process}

\begin{figure}
 \centering
 \includegraphics[width=\textwidth]{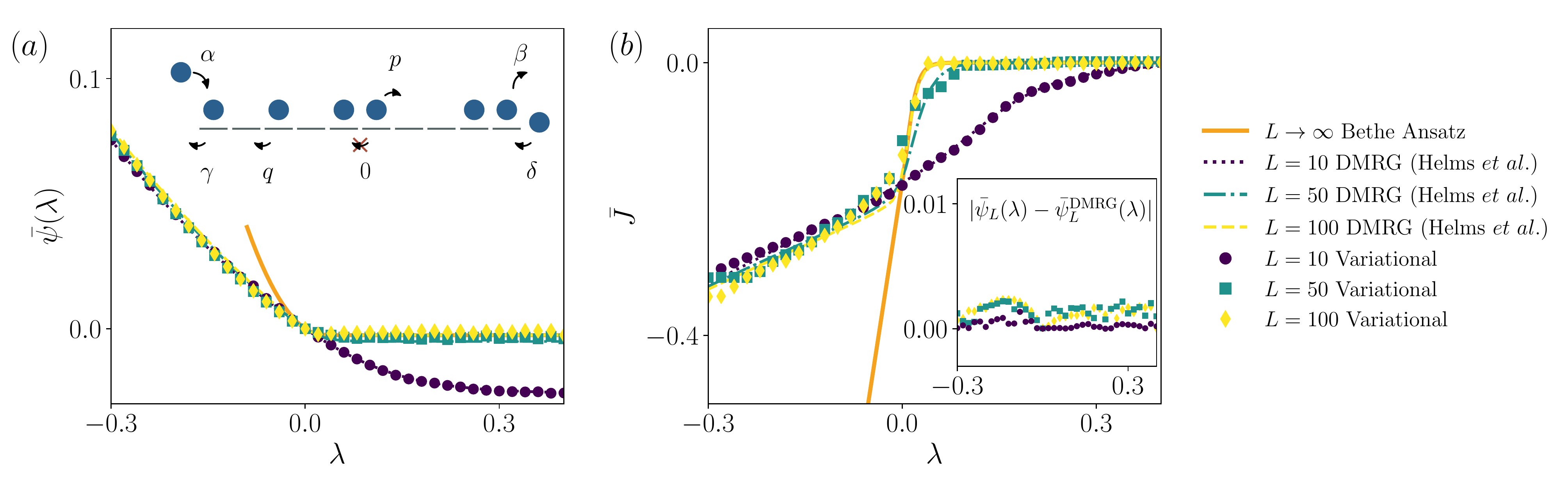}
 \caption{The 1D ASEP with lattice size $L=[10,50,100]$. The variational results (dots) in different lattice size $L$ and DMRG results (dashed lines) of (a) the SCGF scaled by $L$, (b) the average current per site. The variational results of the average current shown in (b) are obtained by numerically computing the derivative of the convex envelope of the corresponding SCGF from (a). The convex envelope is obtained by performing the Legendre-Fenchel twice. The insert of (c) shows the absolute error between the variational results and the DMRG results. The parameters used throughout this example are $p=0.1$, $q=0.9$, $\alpha=0.5$, $\beta=0.5$, $\gamma=0.5$, $\delta=0.5$. For the parameters used in the neural network, the dimension of the hidden layer is $20$, and the number of $L_i$ in \eqref{eq:mlp_1} is $3$. The DMRG results are adapted from \citet{helms2019dynamical}.}
 \label{fig:asep_stat}
\end{figure}

\begin{figure}
 \centering
 \includegraphics[width=\textwidth]{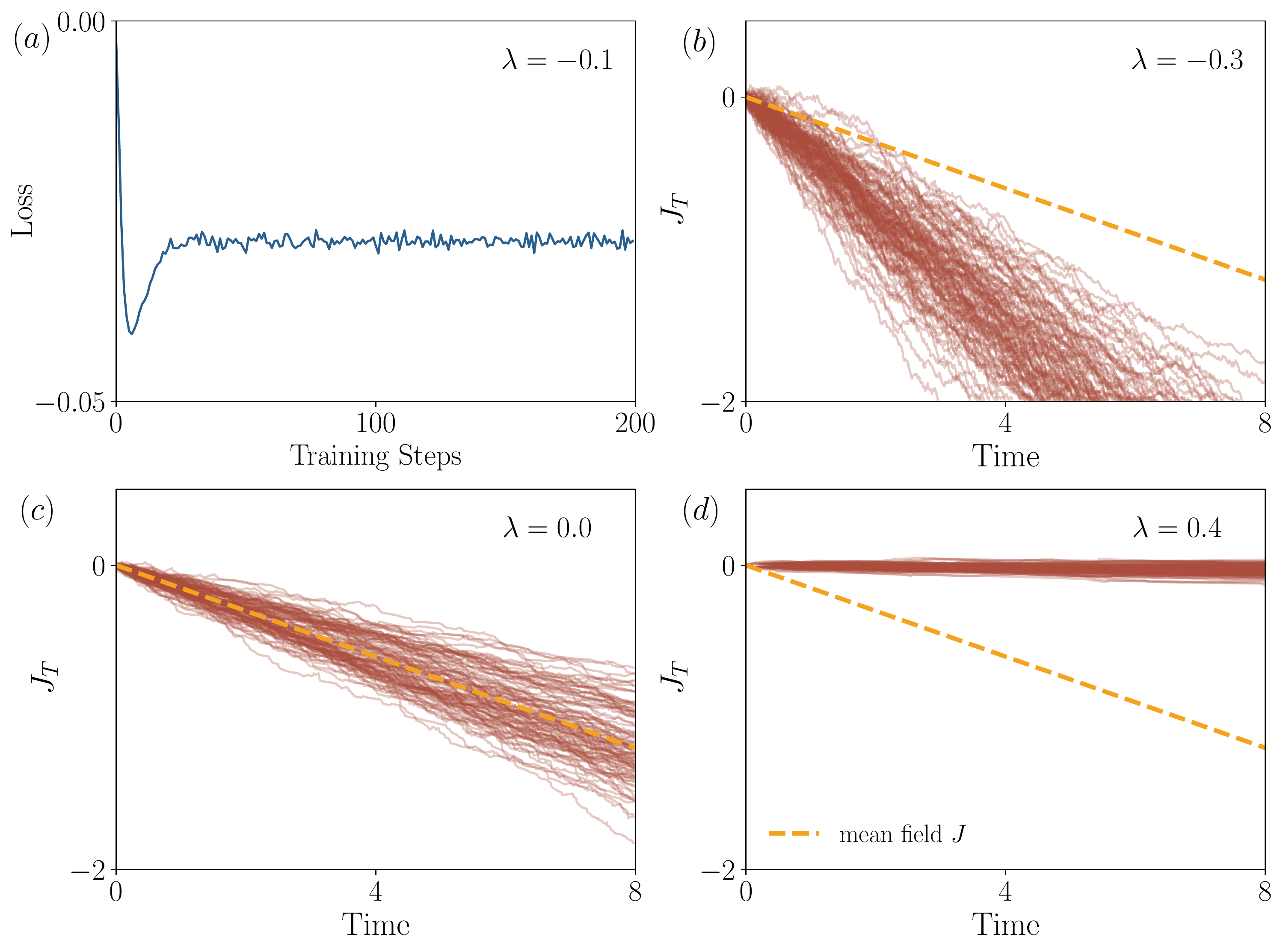}
 \caption{(a) A typical training process for $\lambda=-0.1$. (b)-(d) Typical trajectories of the current in the biased ensemble. The yellow line depicts the average current from the mean field theory. It is clearly that the control force changes the current for different $\lambda$.}
 \label{fig:asep_traj}
\end{figure}

While the variational formulation of our approach gives us a natural metric to compare the quality of the results, without an analytical solution (and in the absence of competitive numerical techniques) we cannot guarantee that the algorithm finds a true maximizer of~\eqref{eq:variational_formula}.
To test the accuracy of our algorithm and whether it can be applied to broad classes of systems, we study the current fluctuations in the asymmetric simple exclusion process (ASEP). 
Exclusion processes are a paradigmatic model for nonequilibrium transport phenomena and their complex dynamical fluctuations~\cite{derrida1998exactly, chou2011non}. 
Although this model has been studied extensively and many analytical results have been derived \cite{lazarescu_exact_2015}, it is still computationally challenging due to the high dimensionality of its state space and its stochastic dynamics.
Together, these properties make this model ideal for testing new computational methods. 

Recently, tensor network algorithms including the matrix product state (MPS) and density-matrix renormalization group (DMRG) have been adapted to study rare fluctuations in lattice models such as the ASEP and the kinetically constrained models \cite{banuls2019using, helms2019dynamical, helms2020dynamical}. 
Here we compute the current SCGF in the ASEP with open boundaries and compare with previous numerical results obtained through DMRG by \citet{helms2019dynamical}.
Consider a 1-dimensional lattice with $L$ sites. 
Each site can be occupied by a hard-core particle, which randomly jumps to its neighboring right or left site with continuous-time transition rates $p$ and $q$, respectively. 
Hops to occupied sites are excluded from the dynamics. 
A mass reservoir is coupled to the system at each boundary; the effective chemical potential difference across the lattice is dictated by the rates $\alpha,\beta,\gamma$, and $\delta$ (see the insert of Fig.~\ref{fig:asep_stat} (a) for the illustration). 
The dimension of the state space for a lattice with $L$ sites will be $2^L$, and the probability distribution evolves according to a Kolmogorov forward equation $\partial_t P = W P$, where $W$ is the continuous-time transition matrix.

The total current $J_T$ of the ASEP for time $T$ is $\int_0^T j_t dt$ where $j_t$ is the total rightward hops minus the total leftward hops per unit time.
The empirical total current rate, $j_T$, satisfies a large deviation principle, which is the function we aim to compute. 
The SCGF, similarly, can be computed through a variational formula:
\begin{equation}
 \label{eq:variational_formula_jump}
 \psi(\lambda) = \sup_u \lim_{T\rightarrow\infty}\left\{\lambda j_T[u] -\frac{1}{T}\int_0^T W_u(x^u_{t^-},x^u_{t^+})\left[\frac{W(x^u_{t^-},x^u_{t^+})}{W_u(x^u_{t^-},x^u_{t^+})} - 1 - \log\frac{W(x^u_{t^-},x^u_{t^+})}{W_u(x^u_{t^-},x^u_{t^+})}\right] dt \right\}
\end{equation}
where a different transition matrix $W_u$ is optimized. 
Eq. \eqref{eq:variational_formula_jump} is different from \eqref{eq:variational_formula} because the Radon-Nikodym derivative for Markov jump processes has a different form than the one for SDEs \cite{segall1975radon}. 
We review the derivation in Appendix~\ref{sc:scgf_jump}.

Denote $S_i=1$ when the $i$th site of the lattice is occupied by a particle and $S_i=0$ otherwise. Practically, for the $i$th site, the modified transition rate $[q_u(i), p_u(i)] = [q + \delta q_u(i), p + \delta p_u(i)]$ when $S_i=1$ and $[q_u(i), p_u(i)]  = 0$ when $S_i=0$. 
The rate modification $[\delta q_u(i), \delta p_u(i)]$ is determined by a neural network which input is $[S_{i-m}, S_{i-m+1},\cdots, S_i, \cdots, S_{i+m}]$. 
The parameter $m$ plays the role of a cutoff distance, determining the range of neighboring information the neural network can gather. 
In this example, force equivariance does not arise because the system is one-dimensional, and the input of $f_e(\cdot)$ becomes $x_j - x_i$ where $x_j$ is the location of a particle within the cutoff distance.
Throughout, we used $m=10$, meaning the input can be both negative or positive depending on the location of the neighboring particle.

To compute the SCGF we use automatic differentiation to backpropagate through the dynamics. 
The results are compared with the asymptotic analytical results in the limit of $L\rightarrow\infty$ derived from the Bethe ansatz \cite{lazarescu_exact_2015}, as well as the previous numerical results by the DMRG algorithm from \citet{helms2019dynamical}, shown in Fig.~\ref{fig:asep_stat}-\ref{fig:asep_traj}. 
The DMRG algorithm with sufficient bond-dimension is nearly exact for this model, which we verified by comparing with the numerically exact spectral solution for $L=10$ (data not shown). 
Fig.~\ref{fig:asep_stat} shows that our method is very accurate for both the SCGF and the cumulants obtained from this cumulant generating functions, and the absolute errors are all less than the order of $10^{-2}$.

\subsection{Dynamical phase transitions in Active Brownian particle systems}

\begin{figure}
 \centering
 \includegraphics[width=\textwidth]{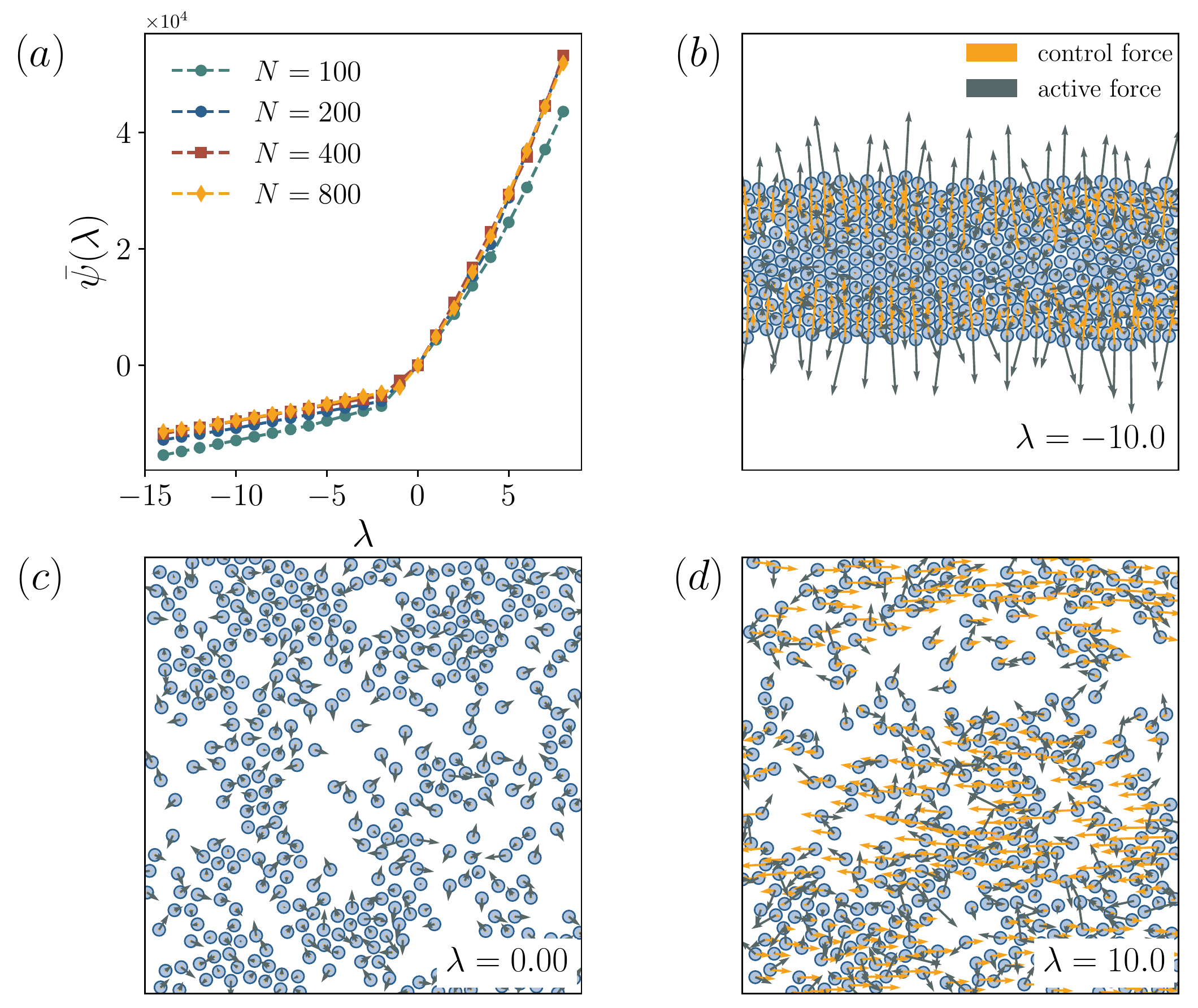}
 \caption{(a) The SCGF of entropy production in the ABP system, with particle number $N=[200,400,800]$. The data suggest that as $N\rightarrow\infty$ there will be a singularity at $\lambda=0$, indicating a first-order dynamical phase transition. The neural network used here is \eqref{eq:nn} with hidden layer dimension equal to $50$ and number of $L_i$ in \eqref{eq:mlp_1} is $3$. (b)-(d) Typical
 snapshots of ABPs under different biasing field $\lambda$ ($N=400$). The red and grey arrows depict the magnitude and direction of the control force and active force $\pmb{b}^{(i)}_t$ for each particle, respectively.}
 \label{fig:abp}
\end{figure}

Recent advances on active matter, such as active Brownian particles (ABPs), active nematics, and driven polymer networks, have articulated a relation between macroscopic pattern formation and energy dissipation~\cite{lamtyugina_thermo_2021, bowick2022symmetry, grandpre_entropy_2021, fodor_dissipation_2020, fodor_how_2019, giomi_defect_2013, narayan_long_2007,nguyen2021organization}. One challenge of characterizing macroscopic pattern formation that results from rare collective fluctuations, is that it requires large system sizes and high densities, making numerical sampling extremely difficult. 
Here we examine our algorithm in the context of the active Brownian particle model, in which the $i$th particle evolves as
\begin{equation}
 \begin{aligned}
  \label{eq:abp}
  d\Xb_t^{(i)} & = [- \mu\diff{U(\Xb_t)}{\xb^{(i)}} + v\bb^{(i)}_t]dt + \sqrt{2D_t}d\Wb_t^{(i)},                   \\
  \bb^{(i)}    & = [\cos\phi_t^{(i)}, \sin\phi_t^{(i)}]^\top,\quad d\phi_t^{(i)} = \sqrt{6D_t}d\Wb_t^{\phi^{(i)}}.
 \end{aligned}
\end{equation}
where $U$ is a purely repulsive WCA potential, $\xb^{(i)}$ denotes the position of the $i$th particle, and $\bb^{(i)}$ denotes the direction of its active velocity.
Throughout we consider systems with particle density $\rho=N/L^2=0.6$, and $v=100$ which is within clustered phase as a result of motility-induced phase separation (MIPS)~\cite{hagan_structure_2013}. 
The non-conservative self-propulsion term $v\bb^{(i)}$ represents the dissipative ``active'' force.
In the ABP model, $\bb_t^{(i)}$ are unit vectors which rotate diffusively and $v$ is the magnitude of the active force. Here, $\Wb_t^{(i)}$ and $\Wb_t^{\phi^{(i)}}$ are independent standard Wiener processes.
The phenomenology of motility induced phase separation is relatively well-understood compared with other nonequilibrium phase transitions: for sufficiently large P\'{e}clet number and density, a macroscopic aggregate of particles forms\cite{hagan_structure_2013}. 
\citet{grandpre_entropy_2021} demonstrated that this transition correlates with the average entropy production
\begin{equation}
 \label{eq:ep}
 s = \frac{1}{NT}\sum_{i=1}^N\int_0^T v\bb^{(i)}_t D^{-1}_t \circ d\Xb_t^{(i)}.
\end{equation}
Mechanistically, when the system enters the phase separated state, much of the directional motion ceases, leading to a drop in the average entropy production compared to an unclustered trajectory.

Previously, we have shown that by using a neural network ansatz, it is possible to compute the many-body control force and the SCGF for systems with up to 200 particles. Here, by using the formalism proposed in Section \ref{sc:algo}, we can employ \emph{transfer learning}.
That is, we train the neural network in a small system (e.g., $N=200$), because the GNN representation is naturally independent of the total size of the system, we can use the trained as an initial condition for larger systems. 
In Appendix~\ref{sc:nn_gnn}, we show that the GNN formulation of the present work outperforms an ``uninformed'' neural network. 

We show the results for the entropy production rate function in Fig.~\ref{fig:abp}. 
In Fig.~\ref{fig:abp}(a) we see clear convergence as $N$ increases, and a singularity at around $\lambda=-1$ indicates a first-order dynamical phase transition, consistent with previous studies\cite{grandpre_entropy_2021, yan2022learning}. 
For positive $\lambda$, the control force induces a collective flow to increase the average entropy production rate as shown in Fig.~\ref{fig:abp} (d). 
When $\lambda$ is negative, below the transition point, particles cluster, reducing mobility and subsequently the entropy production.

\begin{figure}
 \centering
 \includegraphics[width=.8\textwidth]{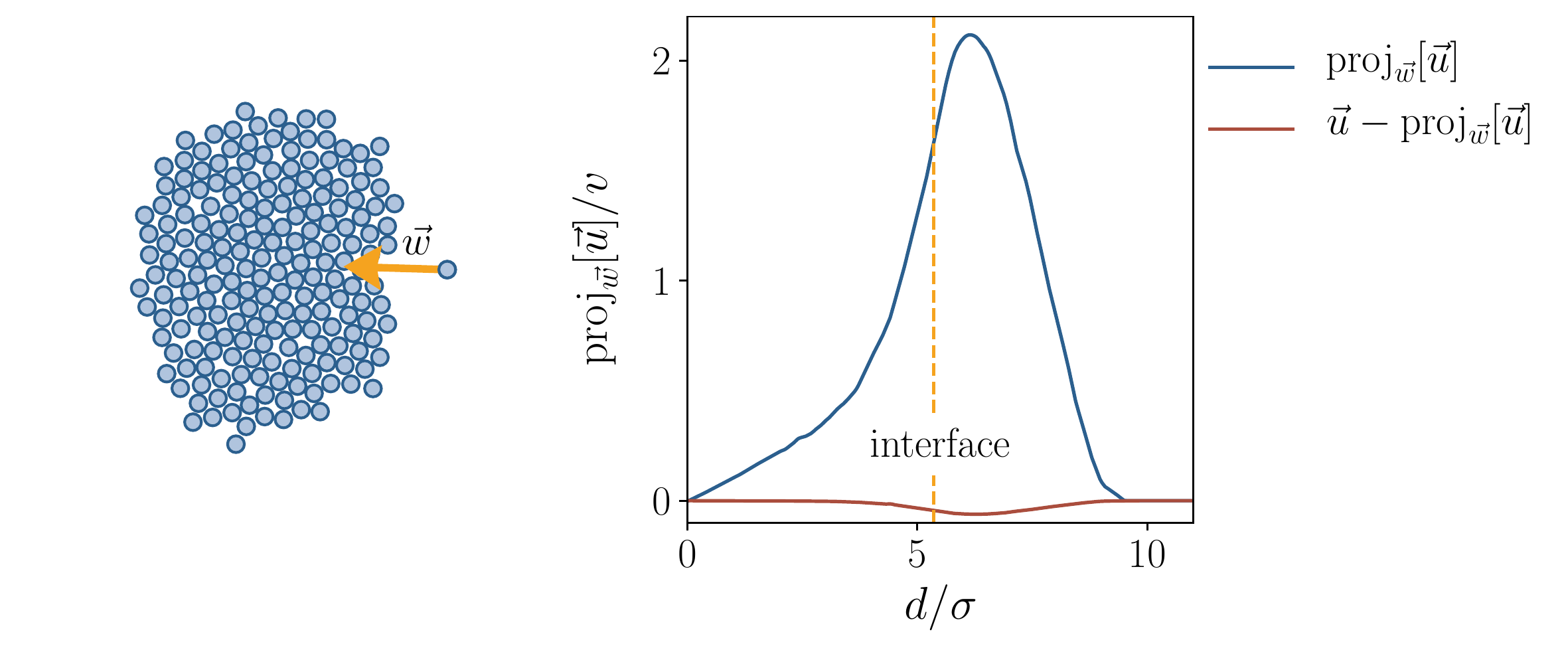}
 \caption{The learned control force of one single particle as it gradually approaching an established cluster ($\lambda=-10$, $\rho=0.1$, $v=100$). We plot the magnitude of the control force in two orthogonal directions, perpendicular to and along the vector $\vec{w}$, which connects the particle and the cluster's center of mass. The force is in the unit of the velocity $v$, and the distance is in the unit of the particle size $\sigma$. 
 The yellow dashed line represents the position of the interface of the cluster. As expected, the control force is small when a particle is inside the cluster, and reaches its maximum when closing to the cluster.}
 \label{fig:gnn_force}
\end{figure}

One conceptual advantage of the optimal control formalism is that we can directly access the control forces that realize the target rare event. 
To better understand the physical forces that drive low entropy production clustered dynamics, we plot the learned control force of one single particle as a function of the distance between it and the center of a cluster (Fig.~\ref{fig:gnn_force}). 
The results show that for particles deeply inside the cluster,r the control force is nearly zero, as it is stabilized by the active forces. 
The force gradually increases near the boundary of the cluster surface and reaches a  maximum when a particle is just outside the cluster.
The force vanishes at the cutoff distance $d_{\rm c}=4\sigma$ by construction, but its relatively long range emphasizes the effective attraction that drives clustering. 

\section{Conclusion and future work}

Many of neural networks used in practice, such as convolutional neural networks and transformers~\cite{vaswani_attention_2017} have become popular because of their success on ``big data'' tasks, computer vision and natural language processing.
Problems in scientific computing demand a distinct class of function representations. 
The physics-informed graph neural networks we have developed in this work not only provide a representation that is specially tailored to the symmetries and properties of the systems we consider, they also are naturally invariant to the extent of the system. 
The ansatz that we use builds in translational invariance and rotational equivariance.
Such a constraint limits the function class, likely accelerating learning, as well.  
Together, these properties manifest in scalability: we can optimize the control forces on small systems where data collection is inexpensive computationally, and then deploy and refine them on systems large enough to explore collective phenomena. 

Recently, tensor network algorithms including MPS and DMRG have shown promise in dynamical large deviation calculations, yet these approaches are much more tailored to lattice models with MJP dynamics.
The ansatz used for DMRG uses much a longer effective range of correlations when the bond-dimension is high. 
The results we obtain for the ASEP model show our approach achieves good performance compared with DMRG, providing an alternative when DMRG is ill-suited to the system in question.

In this work, we consider only homogeneous systems in which there is only one type of particle.
Extending the current framework to heterogeneous systems, for example, multi-component active materials or molecular glass formers, remains a significant future challenge.
Furthermore, since our physics-informed GNNs provide an expressive representation of potentially any physical force, we could apply these networks to many other problems, including nonequilibrium response calculations.

\begin{acknowledgments}
 The authors thank Todd Gingrich for several helpful discussions. We also thank Phillip Helms for generously sharing the DMRG data used in Fig.~\ref{fig:asep_stat}. This work was partially supported by the Terman faculty fellowship. 
\end{acknowledgments}

\bibliography{refs}

\appendix

\section{Variational formula of SCGF for Markov jump processes}
\label{sc:scgf_jump}
The Radon-Nikodym derivative \eqref{eq:girsanov_sde} is a result of the Girsanov theorem \cite{oksendal_stochastic_1992}. 
For Markov jump processes, the change of path measure is obtained via a similar procedure, cf.~\citet{segall1975radon}. 
For simplicity, here we also review an alternative proof by using the contraction principle of the large deviation theory~\cite{chetrite_variational_2015}.

Consider a general Markov jump process on the state space $\mathcal{X}$, and $x, y\in\mathcal{X}$. Let $W(x,y)$ be the continuous-time transition rate from $x$ to $y$. 
Level 2.5 large deviation theory~\cite{chetrite_variational_2015} (cf. Appendix B, therein) states that the joint rate function of the empirical density $\rho(x)$, i.e., the proportion of time that a trajectory stays at $x$, and the empirical flow $q(x,y)$, i.e., time average of the number of jumps from $x$ to $y$, can be explicitly written as
\begin{align}
\label{eq:level_2.5}
 K(\rho, q) & = \sum_{x,y} \left[ q(x,y)\log\frac{q(x,y)}{\rho(x)W(x,y)} - q(x,y) + \rho(x)W(x,y) \right].
\end{align}
From the contraction principle, \eqref{eq:level_2.5} can be obtained by contracting over all control forces $u$, implemented through the transition rate matrix $W_u$,
\begin{equation}
\label{eq:contraction}
    K(\rho, q) = \min_{u}K(\rho_u, q_u) = \sum_{x,y} q_u(x,y)\left[\frac{\rho_u(x)W(x,y)}{q_u(x,y)} - 1 - \log\frac{\rho_u(x)W(x,y)}{q_u(x,y)}\right],
\end{equation}
where $\rho_u$ and $q_u(x,y) = \rho_u(x)W_u(x,y)$ and stationary density and flow associated with the new transition rate matrix $W_u$. 
Simplifying the expression above, we have
\begin{equation}
\label{eq:level_2.5_contrac}
    K(\rho_u, q_u) = \sum_{x,y} \rho_u(x)W_u(x,y)\left[\frac{W(x,y)}{W_u(x,y)} - 1 - \log\frac{W(x,y)}{W_u(x,y)}\right].
\end{equation}

Because the dynamics is ergodic, the ensemble average \eqref{eq:level_2.5_contrac} is equivalent to averaging over trajectories $\{x_t^u\}_{0\leqslant t \leqslant T}$ determined by the rate matrix $W_u$; as a result, we can write
\begin{equation}
    K(\rho_u, q_u) = \lim_{T\rightarrow\infty}\frac{1}{T}\int_0^T W_u(x^u_{t^-},x^u_{t^+})\left[\frac{W(x^u_{t^-},x^u_{t^+})}{W_u(x^u_{t^-},x^u_{t^+})} - 1 - \log\frac{W(x^u_{t^-},x^u_{t^+})}{W_u(x^u_{t^-},x^u_{t^+})}\right] dt.
\end{equation}
The SCGF for the current can be estimated by optimizing the transition rate matrix $W_u$ to perform the contraction in \eqref{eq:contraction},
\begin{align}
    \psi(\lambda) = \sup_u \lim_{T\rightarrow\infty}\left\{\lambda j_T[u] -\frac{1}{T}\int_0^T W_u(x^u_{t^-},x^u_{t^+})\left[\frac{W(x^u_{t^-},x^u_{t^+})}{W_u(x^u_{t^-},x^u_{t^+})} - 1 - \log\frac{W(x^u_{t^-},x^u_{t^+})}{W_u(x^u_{t^-},x^u_{t^+})}\right] dt \right\}.
\end{align}




\section{Comparison with a general purpose Neural Networks}
\label{sc:nn_gnn}
\begin{figure}
 \centering
 \includegraphics[width=.8\textwidth]{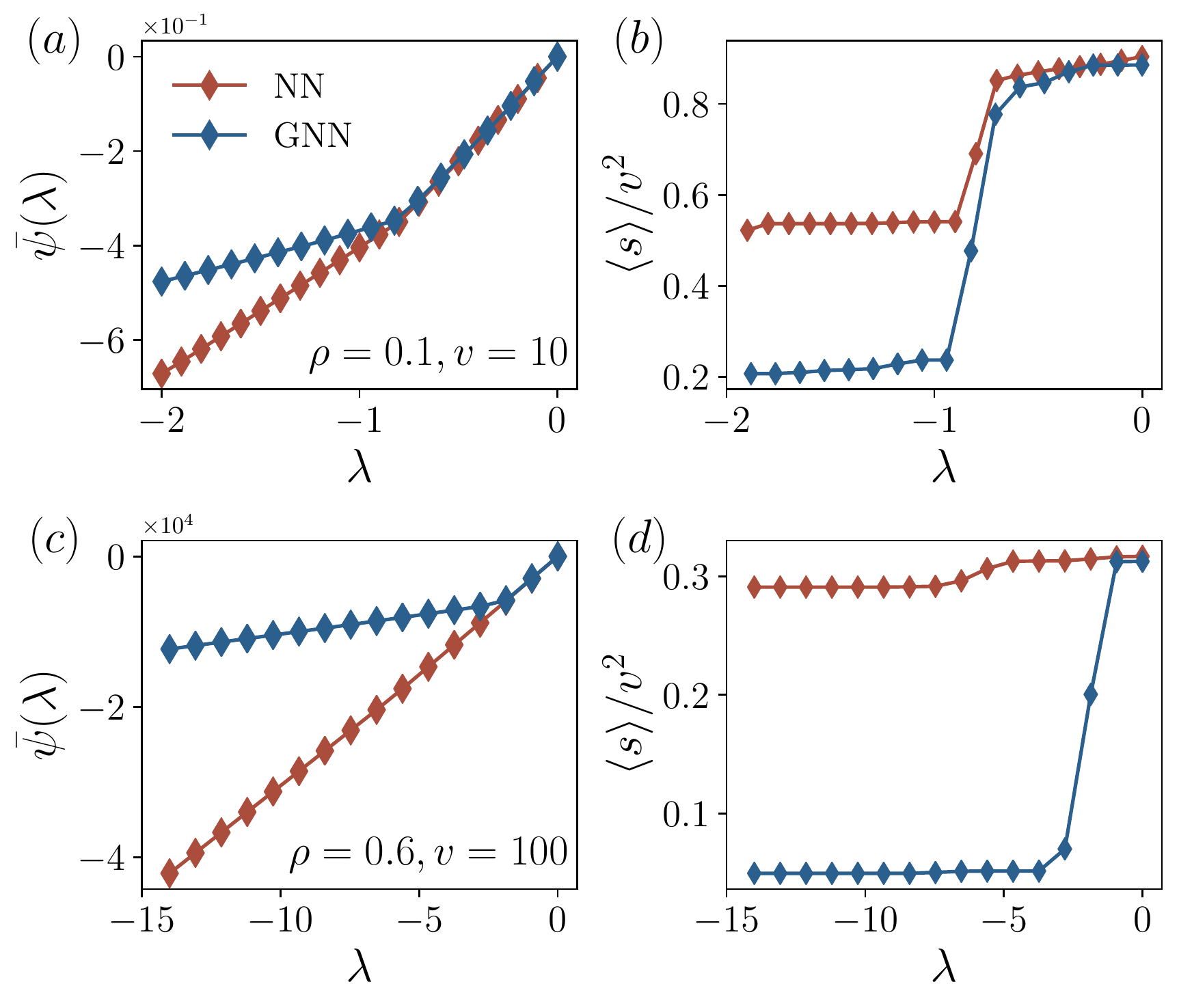}
 \caption{Comparison of the performances of the GNN approach and the NN approach using the ABPs model. We test two set of parameters: (a)-(b) $\rho=0.1$, $v=10$ and (c)-(d) $\rho=0.6$, $v=100$ and plot the SCGF in (a), (c); and the average entropy production per particle in (b), (d). For the GNN approach, we use two $L_i$ (see \eqref{eq:mlp_1}) with hidden layer width equal to $100$. For the NN approach, we use two $L_i$ with hidden layer width equal to $2000$.}
 \label{fig:gnn_vs_nn}
\end{figure}

In previous work \cite{yan2022learning} we used a single neural network to generate a many-body control force for all particles. 
To compare the performance of this general purpose neural network (NN) approach with the graph neural network (GNN) approach presented in this work, we here test two methods in the ABPs model. 
We tested two set of parameters (number density $\rho$ and active force velocity $v$) which lays in different phases in the ABPs model \cite{hagan_structure_2013}. 
The results are shown in Fig.~\ref{fig:gnn_vs_nn}. 
Due to the variational principle we derive, the larger value of $\psi(\lambda)$ must be closer, if not equal, to the true results. In the ABP model, it seems that both method perform equally well for small biasing $\lambda$, but the GNN approach proposed in this work outperforms the previous approach when $\lambda$ becomes more negative. Furthermore, when $v$ is large, the previous approach even fails to capture the nature of the dynamical phase transition, characterized by a singularity of $\psi(\lambda)$ (Fig.~\ref{fig:gnn_vs_nn}(c)-(d)), while the GNN approach still remains robust.

\section{Numerical details of the ABPs model}
In the active Brownian particle model \eqref{eq:abp}, the conservative inter-particle force is modeled as a purely repulsive WCA pair potential $U = \sum_{i\neq j} u(l_{ij})$ which depends on the relative distance between any two particle $i$ and $j$, denoted $l_{ij}$.
Explicitly,
\begin{equation}
 u(l_{ij}) = \left\{\begin{array}{lcl}
  4\epsilon\left[ (\sigma/l_{ij})^{12} - (\sigma/l_{ij})^{6}  \right] + \epsilon, & l_{ij}\leqslant 2^{1/6}\sigma \\
  0, & l_{ij} > 2^{1/6}\sigma,
 \end{array} \right.
\end{equation}
In the simulation, the unit of length is normalized by $\sigma$ and we set $\epsilon=1$. We used periodic boundary condition with box length $L$. The particle number density $\rho$ in Fig.~\ref{fig:abp} is $0.6$ and $v=100$. In Fig.~\ref{fig:gnn_force} we use $\rho=0.1$ and $v=100$. When $\rho$ is high enough, particles aggregate into a band rather than a circular cluster.

\end{document}